\documentstyle{article}
\begin{document}
\section{Introduction}

The Lattice Boltzmann Equation (LBE) was introduced at the turn of the 80's
mainly to cope with two major drawbacks of its ancestor, the Lattice
Gas Cellular Automaton (LGCA) \cite{FHP}.
Ever since, it has undergone a number of refinements and extensions which
have taken it to the point where it can successfully compute a number
of non trivial flows, ranging from homogeneous incompressible turbulence 
to multiphase flows in porous geometries \cite{PHD,BSV,QSO,QLO}.
Yet, when compared with advanced computational fluid dynamics (CFD)
methods, it is apparent that there's still some way to go before
LBE can achieve full 'engineering status'.
This applies especially to flows in complex geometries where highly
irregular, possibly adaptive, meshes are a must.
At the same time, it is also clear that, being sharply aimed at macroscopic
fluid dynamics, LBE does not share the same degree of physical fidelity of
'true' microscopic methods, such as Direct Simulation Montecarlo (DSMC), let
alone Molecular Dynamics (MD).
The question, central to the 'leit motif' of this Conference, comes 
quite naturally: is LBE a powerful blend of the micro-macro worlds keeping
the best of the two, or rather just an unremarkable hybrid giving up
microfidelity with no corresponding computational returns to
outdo 'conventional' CFD? 
In this paper we shall bring arguments in favour of the former alternative. 

\section{The Lattice Boltzmann equation}

The LBE is a minimal discrete Boltzmann equation reproducing
Navier-Stokes hydrodynamics in the limit of small Knudsen
numbers, i.e. particle mean free path much smaller than 
typical macroscopic variation scales \cite{BSV}.

With reference to the earliest 24-speed, single-energy, four-dimensional FCHC 
(Face Centered HyperCube) lattice defined by the condition \cite{DLF}

$\sum_{k=1,4} c_{ik}^2 = 2 \;\;,\; c_{ik}= (0, \pm 1 )$,

the LBE takes the form of a set of first-order explicit finite difference
equations:

\begin{equation}
f_i (x_k + c_{ik}, t+1) - f_i (x_k,t) = \sum_{j=1}^{24} A_{ij}(f_j -f_j^e)
\end{equation}

where $f_i, i=1,24$ is a set of 24 populations moving along a corresponding set
of discrete speeds $c_{ik}$ and the index $k=1,4$ runs over the spatial dimensions. 

The eq.(1) is naturally interpreted as a multi-relaxation scheme in
which non-equilibrium gradients are brought back to the local
equilibrium $f_j^e$ by scattering events mediated by the matrix $A_{ij}$.
The local equilibrium populations are chosen in the form of a discretized Maxwellian
expanded to second order in the Mach number in order to retain convective effects.

\begin{equation}
f_i^e = \frac{\rho}{24} (1 + 2 c_{ik} u_k + 2 (c_{ik} c_{il} - \frac{1}{2} \delta_{kl}) u_k u_l)
\end{equation}

Under the constraint of point-wise mass and momentum conservations
($\sum_i A_{ij} = \sum_i c_{ik} A_{ij} = 0$) 
the equation (1) describes the motion of a {\it four-dimensional} fluid.
The reason for working in four-dimensions is that no 
{\it single-energy} discrete lattice exists 
in three-dimensions ensuring isotropy of the fourth order tensor
$T_{klmn} = \sum_i c_{ik}c_{il}c_{im}c_{in}$.
Such an isotropy is compelling in order to recover rotational invariance
of the momentum flux tensor at a macroscopic level. 
With these preparations, fluid variables (density,speed and
momentum flux tensor) are defined as follows:

$\rho        = \sum_{i=1}^{24}   f_i$, 
$\rho u_k    = \sum_{i=1}^{24} f_i c_{ik}$, 
$P_{kl} = \sum_{i=1}^{24} f_i c_{i,k}c_{i,l}$

and can be shown to obey the Navier-Stokes equations in the continuum limit.

In the recent time, the FCHC model has been superseded by the so-called
BGK (after Bathnagar, Gross and Krook in continuum
kinetic theory) versions, in which the scattering matrix $A_{ij}$ is replaced by a
diagonal form $-\omega \delta_{ij}$ \cite{BGK}.
Here $\omega$ is the inverse
relaxation time controlling the flow viscosity according to the simple relation

\begin{equation}
\nu \sim (-1/\omega -1/2).
\end{equation}

The resulting numerical scheme is {\it linearly stable} for all positive values of
$\nu$ ($-2 < \omega < 0)$, which means that arbitrarily high Reynolds numbers 
$Re=UL/\nu$, $U$ and $L$ being typical macroscopic flow speed and size
respectively, can be attained by choosing $\omega$ sufficiently close to $-2$.

Current practice shows however that 
{\it non linear instabilities} set in if the viscosity is lowered
below a given threshold $\nu_{NL} > 0$.
This nonlinear viscosity threshold is associated with the violation
of the positivity constraint on the discrete populations ($f_i > 0$) and it is
typically triggered whenever the fluid flow develops sharp gradients comparable
in scale with the lattice pitch $\Delta$.
The result is that, like any other floating-point based method, LBE is exposed to
numerical diseases whenever the lattice is not fine enough to resolve the 
shortest physical scales.

\section{LBE in perspective}

LBE is a minimal hyperbolic superset of the Navier-Stokes equations, where
minimal means: "the least amount of information from velocity space" 
required to recover the correct continuum space-time symmetries 
(Galilean, translational and rotational invariance).
Symbolically, we write $ K = H \bigoplus M $
where $K$ is the kinetic space spanned by the
discrete populations $f_i$, $H$ is the subspace spanned by hydrodynamic fields 
and $M$ the subspace associated with mesoscopic fields. 
Along with the hydrodynamic fields (density, flow, temperature)
LBE also tracks the momentum-flux tensor as a well as a number of
higher-order moments of the distribution function that were dubbed
'ghost-fields' since, although key to secure the correct spatial symmetries,
they do not relate explicitly to any hydrodynamic quantity.
They represent genuinely {\it mesoscopic} information.

The fact that the momentum flux tensor is also carried along, means that
LBE is {\it more} than a plain Navier-Stokes solver.
This property is of no use for the direct simulation of fluid turbulence
but it becomes very valuable when it comes to turbulence modeling since
it permits straightforward and cheap implementations of  stress-tensor
dependent effective viscosities such as those used in algebraic 
turbulence models \cite{SOM,CSD}.
Ghost fields are an inevitable overhead, somehow
the price to pay to formulate hydrodynamics in a fully covariant form.
It should be appreciated that this covariant (read hyperbolic) formulation 
brings about two interesting numerical implications: i) diffusion does not
require second order space derivatives, ii) convection takes place along 
constant streamlines defined by the condition $\Delta x_{ki} = c_{ki} \Delta t$.
Both features prove extremely beneficial for the amenability of
LBE to parallel computing.
Ideally, one would like to draw some further benefit from ghost fields, namely use
the mesoscopic scales to do 'some good' to the hydrodynamic ones.
As they stand, unfortunately ghost fields don't seem to provide any beneficial effect
on the non-linear stability properties of the scheme.
However, they can be manipulated to construct subgrid turbulence models
\cite{BSV} which, regretfully enough, have not been implemented so far.
So, to date, one must conclude that no positive feedback between the mesoscopic
and the macroscopic components of LBE has been put in operation.

In CFD parlance we would classify LBE as a 
{\it
"Strictly synchronous, first order in time, second order
 in space, fully explicit finite difference scheme".
}

Within this wide class of CFD schemes, the main hallmarks of LBE are

\begin{itemize}
 \item Point-wise conservativeness to machine round-off
 \item Unconditional linear stability
 \item Diffusion-dispersion freedom
 \item Finite hyperbolicity
 \item High compute density (operations/grid point)
\end{itemize}

Main pro's are:

\begin{itemize}
 \item Amenability to parallel computing
 \item Easy handling of grossly irregular geometries (e.g., porous media)
 \item Ease of use and physical soundness
\end{itemize}

Counteracting con's

\begin{itemize}
 \item Uniform-grid boundness 
 \item Computational redundance (ghost fields)
\end{itemize}

One may notice that all of these points, both pro's and con's are
technical in nature rather than fundamental.
This reflects the historical focus of LBE on fluid-dynamics,
and-to a good extent- the author's own bias as well.

Happily enough, some researchers have wisely turned
an eye also to the 'fundamental' direction, and worked out ways to
extend the method so as to embed genuinely mesoscopic physics.
Remarkable work along this line has been performed in the
recent past \cite{ROT,LAD,YEO,ORL}
and interesting attempts to 'walk upwards'
the BBGKY hierarchy are just starting to appear \cite{GROSF}.

\section{Success}

As of today, almost ten years after its inception, 
the body of LBE calculations is simply too wide to be
covered by any single comprehensive paper. 
Best approximation, and yet already partially outdated, is the recent review
paper by Qian et al \cite{QSO}.

In a way, this is probably the best sign of success.

Many of these calculations may have been successfully
performed by other CFD methods as well, but probably, not all
of them. It looks like some multiphase flows calculations
appear to come by much more easily-if not more efficiently-using 
LBE rather than 'conventional' grid-based techniques.
Particularly enticing to the physicist's taste is the fact most
of the new physics can be encoded into suitable generalizations
of local equilibria and/or phenomenological source terms reflecting the 
assumptions on the underlying microscopic physics.
Whether there is more to this explicit link between physics and numerics
than esthetical satisfaction is difficult to say; for sure, it is an extremely
pleasing feature of the method.
An indisputable fact is the success of LBE applications
on (virtually any kind of) parallel computers.
As a recent first-hand example, our group in Rome is currently running turbulent 
channel flow LBE simulations at sustained speeds in excess of $10$ Gflops using
massively parallel SIMD machines \cite{FVLBE}.
And even at the industrial level, LBE high amenability to parallel
computing has opened the way to complex flow calculations such as those performed
at Shell Research.
%\cite{EGG}.

\section{Open problems and future developments}

Two major stumbling blocks have been holding back 
engineering and physics LBE applications:

\begin{itemize} 
\item Uniform-grid boundness
\item Non linear stability of thermohydrodynamic schemes
\end{itemize}

While the former limitation has been basically overcome in the
recent years, the latter is still with us.
Let us comment in some more detail.

\subsection{Uniform Grid}
To the best of the author's knowledge, to date, nobody knows how to extend synchronous
LBE schemes of the form given by eq. (1) to generalized coordinates
and/or irregular mesh distributions.
This is a major stumbling block, as it rules out
a host of important real-life engineering applications.

The problem has been (partially) circumvented by marrying
LBE with standard Finite-Volume and, more recently, 
Finite-Difference techniques \cite{FVLBE}.

The basic idea is to focus on the continuum version of LBE, namely
\begin{equation}
[ \partial_t + c_{ik} \partial_k ] f_i (x_k, t) = \sum_{j=1}^{24} A_{ij}(f_j -f_j^e)
\end{equation}

and recognize that this is nothing but a set of hyperbolic partial differential equations.
which can handled by any of the commonly available discretization techniques.

Manifestly, by adopting a generic time marching and space discretization
scheme, the synchronization between discrete speeds and the spatial stencil 
is generally spoiled ($\Delta x_{ik} \neq c_{ik} \Delta t$). 
So, one may refer to these extensions as to {\it asynchronous} LBE (ALBE).
This is precisely where the geometrical flexiblity comes from, since
the spatial stencil is now set free from the symmetry requirements imposed
to the discrete speeds. 
The price for geometrical freedom is the need to interpolate between the 
particle positions generated by the discrete speeds and the sites of the 
spatial grid.
Besides sheer computational extra costs (easily offset by conspicuous savings
in the number of grid points), this interpolation may introduce
numerical diffusion effects which are still awaiting for a definitive assessment.

All in all, it is fair to say that much of the initial gap between LBE and
advanced CFD in terms of handling complex geometries is going to close up
in the near future.

\subsection{Non-linear stability}  

When discussing non-linear stability, it is useful to
bear in mind that the discrete grid sets 
an intrinsic lower bound to the flow viscosity (upper bound on Reynolds number) 
that can be used in the numerical simulation of the Navier-Stokes equations.
This lower bound is given by:

$\nu_{\Delta} = u_{\Delta} \Delta$

where $u_{\Delta}$ is the velocity field at the grid scale $\Delta$.

This limit, a real sort of {\it Numerical Uncertainity Principle}
(NUP) cannot be violated by any {\it direct} simulation method
of the Navier-Stokes equations 
without incurring numerical bankruptcy or, worse, fake physical results.
Isothermal LBE calculations can systematically
be brought down to viscosities pretty close to the NUP value, with 
most empirical evidences for a non-linear
viscosity threshold within a $50$ percent higher than the NUP value.

For 'hot', thermodynamic calculation the situation is much less clear.
According to some researchers, the stability threshold undergoes a
severe deterioration which makes LBE's totally unsuited to
thermohydrodynamic computations \cite{MGA}.
That may indeed be the case since thermal LBE schemes are highly remnant
of high-order finite differences stencils whose exposure to numerical
instabilities is pretty well known.
However, a lot more quantitative work is required, before any conclusive 
statement can be drawn in this respect.

In particular two promising ideas are worth being mentioned here.

The first consists of using higher-order expansions in the Mach number of local
equilibria \cite{YUCHEN}. 
A close inspection of these schemes
reveals the appearance of {\it negative definite} equilibrium populations; 
at a first glance, this may sound like complete nonsense, but,
with a little bit of optimism, one might observe that negative populations do not hurt
stability as long as they stay negative all the time. In this case, one may just
reinterpret them as 'holes' or 'antiparticles' and hope their presence help decrease
the non-linear viscosity threshold by partially relaxing the constraint that {\it all}
populations be positive. In fact, a blend of particles and antiparticles could
prove more stable; the compelling condition being that they don't transform one into another,
i.e. populations should not change sign under any flow condition.
A second, more solid, possibility is to use full Maxwellian equilibria rather than
Mach number expansions \cite{WAGNER}.
The obvious advantage is that all equilibrium populations are
positive by definition. The price is computational overhead since the Lagrangian
multipliers of the local maxwellian cannot be assigned a priori but
need be recomputed on the fly, thus implying a small non-linear algebraic
problem at each lattice site and each time step.
Alternatively, a pre-computed table look-up could also be employed.

Whether or not these ideas will prove successfull remains to be seen; for sure
they look promising and deserve thorough exploration before discrete-speed
models are ruled out for thermohydrodynamic applications. 

\section{Conclusions}

As mentioned in the Introduction, LBE was generated as an off-spring
of lattice Gas Cellular Automata (LGCA).
Most of the excitement behind LGCA was driven by the "Grand-Dream":

{\it
LGCA : Turbulence = Ising Model : Phase Transitions
}

Ten years later, all reasonable indications are that the "Grand-Dream"
has turned into a "Grande-Illusion" (but, who knows for the future?).

LBE was born on a much less ambitious footing; just provide a useful
tool to investigate fluid dynamics and, maybe mesoscopic phenomena,
on parallel machines. And in that respect, it appears hard to deny
that, even though much remains to be done, the method has indeed
lived up to the initial expectations.
As for the future, competition with macroscopic CFD does not promise
any spectacular breakthrough; here LBE should probably be regarded as "yet an other
finite difference scheme", very easy to use and implement, very rewarding on
parallel computers, but with no compelling edge over existing methods.
The situation may turn for the best on two conditions: i) relax uniform grid constraint 
and ii) make intelligent use of mesoscopic information to access higher Reynolds number
flows. The former goal is partially achieved, the latter is currently left unexplored.

Use of LBE for truly mesoscopic fluid dynamics problems appears to be
the "golden avenue" for the future; whenever a reliable mesoscopic LBE scheme 
can be put at work, the resulting computational gain over existing microscopic 
methods (Direct Simulation Monte Carlo, Molecular Dynamics) will not score 
factors but (several) orders of magnitude.

\section{Acknowledgements}

Enlightening discussions with B. Alder, M. Mareschal, B. Hoover and H. Posch are
kindly acknowledged.


\begin{thebibliography}{9}
\bibitem{FHP} 
U. Frisch, B. Hasslacher, Y. Pomeau,
Lattice gas Cellular Automata for the Navier-Stokes equations,
{\em Phys. Rev. Lett.\/} {\bf 56} (1986) 1505

\bibitem{PHD} 
G. Doolen editor,
Lattice Gas Methods for PDE's, 
{\em Physica D\/} {\bf 47, 1-2} (1991) 

\bibitem{BSV} 
R. Benzi, S. Succi, M. Vergassola, 
Theory and Application of the Lattice Boltzmann Equation,
{\em Physics Reports\/} {\bf 222(3)} (1992) 147.

\bibitem{QSO} 
Y.H. Qian, S. Succi, S. Orszag, 
Recent advances in Lattice Boltzmann Computing,
{\em Ann. Rev. Comp. Phys.\/} {\bf 3 } (1995) 195.

\bibitem{QLO} 
Y. Qian, J. Lebowitz, S. Orszag ed.s, 
Special issue on lattice Gas,
{\em J. Stat. Phys.\/} {\bf 81,1-2 } (1995).

\bibitem{DLF} 
D. D'Humieres, P. Lallemand, U. Frisch,
Lattice gas models for three-dimensional hydrodynamics,
{\em Europhys. Lett.\/} {\bf 2 } (1986) 291.

\bibitem{BGK} 
Shiyi Chen, Hudong Chen, Daniel Martinez and William Matthaeus,
Lattice Boltzmann Model for Simulation of Magnetohydrodynamics,
{\em   Phys. Rev. Lett.\/} {\bf 67} (1991) 3776.

Y.H. Qian, D. d'Humieres, P. Lallemand, 
Lattice BGK models for the Navier-Stokes equations,
{\em Europhys. Lett.\/} {\bf 17} (1992) 479.

\bibitem{SOM} 
J. Somers, P. Rem, 
Flow computations with lattice gas,
{\em App. Sci. Res.\/} {\bf 48 } (1995) 391.

\bibitem{CSD}
S. Chen, Z. Wang, X. Shan, G. Doolen, 
Lattice Boltzmann computational fluid dynamics in three dimensions,
{\em J. Sta. Phys.\/} {\bf 68 3-4 } (1992) 379.

\bibitem{MGA}
F. Alexander, S. Chen, J. Sterling,
Lattice gas thermohydrodynamic models,
{\em Phys. Rev. E\/} {\bf 47} (1993) 2249, and

G. Mc Namara, A. Garcia, B. Alder, 
Stabilization of thermal lattice Boltzmann models
{\em J. Stat. Phys.\/} {\bf 81,1-2 } (1995) 395.

\bibitem{ROT} 
A. Gustensen, D. Rothman, S. Zaleski, G. Zanetti,
Lattice Boltzmann models of immiscible fluids,
{\em Phys. Rev. A\/} {\bf 43, 8} (1991) 4320, and

E. Flekkoy, D. Rothman,
Fluctuating fluid interfaces,
{\em Phys. Rev. Lett.\/} {\bf 75} (1995) 260.

\bibitem{LAD} 
T. Ladd,
Short-time motion of colloidal particles: numerical simulation via
fluctuating hydrodynamics,
{\em Phys. Rev. Lett.\/} {\bf 74} (1995) 318, and

T. Ladd, 
Numerical simulation of particulate suspensions via a discretized
Boltzmann equation: Part I, Theoretical Foundations
{\em J. Fluid Mech.\/} {\bf 271} (1994)

\bibitem{YEO} 
M. Swift, W. Osborn, E. Orlandini, J. Yeomans,
Lattice Boltzmann simulations of non-ideal fluids,
{\em Phys. Rev. Lett.\/} {\bf 75} (1995) 830.

\bibitem{ORL} 
E. Orlandini
{\em these proceedings\/}

\bibitem{GROSF} 
P. Grosfils,
{\em private communication\/}

\bibitem{EGG} 
K. Eggers and J. Somers, {\em Int. J. Heat and Fluid Flow\/},
in press
\bibitem{FVLBE} 
F. Nannelli, S. Succi,
Lattice Boltzmann equation on irregular lattices,
{\em J. Stat. Phys.\/} {\bf 68, 3} (1992), and

G. Amati, S. Succi, R. Benzi,
Turbulent channel flow simulations with a coarse-grained extension of the
Lattice Boltzmann method
{\em Fluid Dyn. Res.\/}, in press (1996)

N. Cao, S. Chen, S. Jin, D. Martinez, 
Physical symmetry and lattice symmetry,
Los Alamos National Lab preprint, 1996

\bibitem{YUCHEN} 
Y. Chen
Lattice BGK method for fluid dynamics,
{\em Ph.D. Thesis, Quantum Eng. Dept, Univ. of Tokyo\/} {\bf} (1995) 

\bibitem{WAGNER} 
A. Wagner,
{\em private communication\/}

\end{thebibliography}
\end{document}